\documentclass[aps,prd,preprint,groupedaddress,showpacs]{revtex4}
\usepackage{amsmath}
\usepackage{amssymb}
\newcommand{\rp}{r_{+}}
\newcommand{\Mi}{M_{\infty}}

\begin{document}


\title{Charged black holes in quadratic gravity}


\author{Jerzy Matyjasek}
\email{matyjase@tytan.umcs.lublin.pl, jurek@kft.umcs.lublin.pl}
\author{Dariusz Tryniecki}
\affiliation{Institute of Physics, Maria Curie-Sk\l odowska University\\
pl. Marii Curie-Sk\l odowskiej 1, 20-031 Lublin, Poland}


\date{\today}

\begin{abstract}
Iterative solutions to fourth-order gravity describing static and
electrically charged black holes are constructed. Obtained solutions
are parametrized by two integration constants which are related to the
electric charge and the exact location of the event horizon. Special
emphasis is put on the extremal black holes. It is explicitly
demonstrated that in the extremal limit, the exact location of the
(degenerate) event horizon is given by $\rp\,=\,|e|.$ Similarly to the
classical Reissner-Nordstr\"om solution, the near-horizon geometry of
the charged black holes in quadratic gravity, when expanded into the
whole manifold, is simply that of Bertotti and Robinson. Similar
considerations have been carried out for the boundary conditions of
second type which employ the electric charge and the  mass of the
system as seen by a distant observer. The relations between results
obtained within the framework of each method are briefly discussed.
\end{abstract}


\maketitle


\section{Introduction}
According to our present understanding applicability of the
conventional Einstein-Hilbert equations is limited to curvatures
significantly less than the Planck scale and should be considered as
the first approximation to a more fundamental theory. And although it
is not clear how to construct this more fundamental theory, it seems
reasonable to address the question of its possible low-energy impact.
In the quest for imprints of the quantum gravity effects in the
classical regime especially helpful observation is that regardless of
the formulation of the fundamental theory its low energy effective
action should consist of the classical gravity supplemented by
covariant higher curvature terms and higher derivative terms involving
other physical fields. The gravitational part of the total action can
be written therefore as
\begin{equation}
I_{G}\,=\,I_{0}\,+\,I_{1}\,+\,I_{2}\,+\,... ,
\end{equation}
where $I_{0}$ is the Einstein-Hilbert action and $I_{i}$ 
(for $i \geq1$) denotes combination of operators of dimension $2i+2$ with
numerical coefficients depending on a type of the theory.

Among the various modifications of general relativity proposed so far
the prominent role is played by Lagrangians describing the quadratic
gravity. The interest in the theories of this very type is motivated
by the fact that such theories appear, as expected, in low energy
limit of string theory and in the attempts to construct renormalizable
theory of gravity coupled to matter~\cite{Utiyama,Stelle1,Stelle2}. It
should be also remembered that theories of this kind are almost as old
as the general relativity itself and as the general relativity have a
noble parentage~\cite{Weyl1,Weyl2,Pauli,sirArthur}.

Higher derivative terms in the equations of the gravitational field
have very important consequences. One of them consists in the obvious
observation that such equations are presumably hard to solve even in
the cases when Einstein's equations admit solutions expressible in
terms of known special functions. Further, it should be emphasized
that although family of solutions is richer one should be cautious in
selecting physically meaningful solutions~\cite{Simon1}. Naive
acceptance of solutions may lead to interesting behavior of the system
but a wrong physics. It is natural therefore that in order to gain
insight into the nature of the nontrivial problem one has to refer to
approximate methods. Such methods are expected to yield reasonable
solutions to the problem simultaneously selecting the physical ones
\cite{Simon1,Simon2,Parker1}.

For almost two decades perturbative methods have been extensively used
in black hole physics in the context of the back reaction of the
quantized fields (both massive and massless) on the
metric~\cite{Jimmy1,LS1,LS2,Hochberg1,Hochberg2,kocio98,
Jirinek01b,THA00,Jirinek03b}. The (one-loop) approximation to the
stress-energy tensor of the massive quantized fields in a large mass
limit for example, could be regarded as higher curvature term constructed from
the curvature tensor, its contractions and
covariant derivatives~\cite{Kofman1,fz82,fz83,Jirinek00,Jirinek01a}. It should
be noted, however, that the back reaction analyses are, inevitably,
limited to the linearized equations.
Similar considerations in the context of the string inspired
action with  quadratic and quartic terms contructed form the  Riemann tensor
have been carried out in Refs.~\cite{Myersss,Callanss}.

The general perturbative  method has been proposed in a series of
papers~\cite{Campanelli:1994sj, Economou:1994va,Campanelli:1995jt},
and, subsequently, successfully applied in a number of physically
interesting cases, as the de Sitter universe, cosmic strings, charged
black holes, and gravitational waves~\cite{Odylio}. The method is
simple: assuming that the quadratic terms could be considered as small
contribution to the effective stress-energy tensor, which is justified
for small curvatures of spacetime, one can iteratively solve resulting
equations order by order starting with the classical Einstein field
equations. Among various applications the most interesting from our
point of view is the perturbative solution describing static,
spherically-symmetric and electrically charged black hole constructed
by Campanelli, Lousto and Audertsch (CLA) in
Ref.\cite{Campanelli:1995jt}. The calculations have been carried up to
third order with a special emphasis put on the thermodynamical issues.
However, their analysis is somewhat obscured by the choice of the
boundary conditions, and, moreover, propagation of errors in the
metric tensor caused errors in some of the results.

In this paper we shall return to this problem and express resultant
solutions in terms of the electric charge and the {\it exact} location
of the event horizon, $\rp.$ It seems that such a choice is more
natural and reveals simplicity of the extremal configuration. It is
also helpful in the analyses of the near-horizon geometry of the
extremal black hole. Our method is similar to that of
York~\cite{Jimmy1} and Lu and Wise~\cite{Lu} with the different choice
of the boundary conditions.

The paper is organized as follows. In Sec.~\ref{sec2} we introduce
basic equations  and briefly sketch employed method. We choose the
line element in the form propounded by Visser~\cite{Matt1}, which has
proved to be a very useful representation. Since the functional
dependence on the metric tensor  of all terms appearing in equations is
known, we start with the exact location of the event horizon from the
very beginning. It means that the higher order terms do
not contribute to $\rp.$ To express our results in a more familiar form we
also introduce the horizon defined mass. Discussion of the temperature
and entropy of the nonextremal configuration is presented in
Sec.~\ref{sec3}. A careful examination of the extremal configuration
and the role played by the Bertotti-Robinson geometry is given in
Sec.~\ref{sec4}. Moreover, it is shown that the extremal black holes are
characterized by $\rp\,=\,|e|.$ In Sec.~\ref{sec5} we discuss the
problem from the point of view of a distant observer and explicitly
demonstrate relations between both choices of the boundary conditions.
Sec.~\ref{sec6} contains final remarks.

\section{Equations}
             \label{sec2}
The coupled system of the electrodynamics and the quadratic
gravity with the cosmological term set to zero is described by the action
\begin{equation}
      I=\frac{1}{16\pi }I_{G}+I_{em},
                                     \label{actionf}
\end{equation}
where 
\begin{equation}
      I_{G}\,=\,\int g^{1/2}\left( R\,+\,\alpha R^{2}\,+\,\beta
	  R_{ab}R^{ab}\right) d^{4}x,
                                     \label{Sg}
\end{equation}
and 
\begin{equation}
      I_{em}\,=\,-\frac{1}{16\pi }\int g^{1/2} F\, d^{4}x.
                                     \label{Sm}
\end{equation}
Here  $F=F_{ab}F^{ab},$ $F_{ab}\,=\,A_{b,a}-A_{a,b} $ and 
all symbols have their usual meaning. The Kretschmann scalar has been
relegated from the action by means of the Gauss-Bonnet invariant.

Of numerical parameters $\alpha$ and $\beta$ we assume that they are
small and of comparable order otherwise they would lead to
observational effects within our solar system. Their ultimate values
should be determined from observations of light deflection, binary
pulsars, and cosmological
data~\cite{Accioly:2001cc,Odylio,Mijic:1986iv}.
Following~\cite{Campanelli:1995jt} we shall restrict ourselves to
spacetimes of small curvatures, for which conditions
\begin{equation}
|\alpha R|<<1, \hspace{1cm} |\beta R_{ab}| << 1,
\end{equation}
hold. Additional constraints could be obtained from a non-tachyon
conditions, which are closely connected with stability of
solutions~\cite{Whitt:1985}. Demanding the linearized equations to
possess  a real mass~\cite{Steve,Audretsch:1993kp}, one obtains
\begin{equation}
3\alpha \,-\,\beta \,\geq 0, \hspace{1cm}\beta \leq 0.
\end{equation}

To simplify calculations, especially to keep control of order of terms
in complicated series expansions, we shall introduce another
(dimensionless) parameter $\varepsilon$ substituting
$\alpha\,\to\,\varepsilon\,\alpha$ and
$\beta\,\to\,\varepsilon\,\beta.$ We shall put $\varepsilon =1$ in the
final stage of calculations. As the coefficient $\alpha$ does not
appear in the final formulas, introduction of the additional parameter
might appear as an unnecessary complication in the present context.
However, it is really helpful when dealing with more general
Lagrangians, as, for example, these of nonlinear electrodynamics.

Differentiating functionally $S$ with respect to the metric tensor one
obtains the system 
\begin{equation}
     G_{a}^{b}\,-\,\alpha\,\phantom{}^{(1)}H^{b}_{a}\,-\,\beta\,
	 \phantom{}^{(2)}H^{b}_{a}\,=\,8\pi\, T_{a}^{b},
                                     \label{grav-sq}
\end{equation}
where 
\begin{equation}
     \frac{1}{g^{1/2}}\frac{\delta }{\delta g_{ab}}\int g^{1/2}Rd^{4}x\,=\,
     -G^{ab},
                                     \label{eg} 
\end{equation}
\begin{eqnarray}
     \phantom{}^{(1)}H^{ab} &=&\frac{1}{g^{1/2}}\frac{\delta }{\delta g_{ab}}\int
     d^{4}x\,g^{1/2}R^{2}  \nonumber \\
   &=&2R^{;\,ab}-2RR^{ab}+\frac{1}{2}g^{ab}\left( R^{2}-4\Box R\right)
                                     \label{ei}
\end{eqnarray}
and 
\begin{eqnarray}
     \phantom{}^{(2)}H^{ab} &=&\frac{1}{g^{1/2}}\frac{\delta }{\delta g_{ab}}\int
     d^{4}x\,g^{1/2}R_{ab}R^{ab}  \nonumber \\
   &=&R^{;\,ab}-\Box R^{ab}-2R_{cd}R^{cbda}+\frac{1}{2}g^{ab}\left(
     R_{cd}R^{cd}-\Box R\right) .
                                     \label{ej}
\end{eqnarray}

The Faraday tensor $F_{ab}$
satisfies 
\begin{equation}
     {F^{ab}}_{;a}\,=\,0
                                      \label{ned1}   
\end{equation}
and
\begin{equation}
     {^{*}F^{ab}}_{;a}\,=\,0.
                                     \label{ned2}  
\end{equation}
The stress-energy tensor, $T^{ab}$, defined as 
\begin{equation}
     T^{ab}=\frac{2}{g^{1/2}}\frac{\delta }{\delta g_{ab}}S_{em}
                                     \label{tep}
\end{equation}
is therefore given by 
\begin{equation}
     T_{a}^{b}=
     {\displaystyle{1 \over 4\pi }}
     \left( 
     F_{ca}F^{cb}-
     {\displaystyle{1 \over 4}}
     \delta _{a}^{b} F \right).
                                     \label{tep1}
\end{equation}

Let us consider the spherically-symmetric and static geometry.
As is well-known the spacetime metric can be cast in the form 
\begin{equation}
     ds^{2}=-f\left( r\right) e^{2\psi \left( r\right) }dt^{2}+
     \frac{dr^{2}}{f\left( r\right) }+r^{2}\left( d\theta ^{2}+
     \sin ^{2}\theta d\phi ^{2}\right)
                                     \label{el-gen}
\end{equation}
with~\cite{Matt1} 
\begin{equation}
     f\left( r\right) =1-\frac{2m\left( r\right) }{r}.
\end{equation}
The metric has horizons at values of the radial coordinate
satisfying 
\begin{equation}
m(\rp)\,=\,\frac{\rp}{2}.
\end{equation}
In what follows we shall restrict our analyzes to the outermost horizon,
i. e., the one for which
\begin{equation}
m(r) < \frac{r}{2}
\end{equation}
for $r > r_{+}.$

Spherical symmetry places restrictions on the form
of the tensor $F_{ab}$: the only nonvanishing components of the 
Faraday tensor
are connected with the static radial electric and magnetic fields.
Simple integration of the Maxwell equations gives
\begin{equation}
F_{01}\,=\,-\frac{a_{1}}{r^{2}}e^{\psi}
                                  \label{F01}
\end{equation}
and
\begin{equation}
F_{23}\,=\,a_{2}\sin\theta,
\end{equation}
where $a_{1}$ and $a_{2}$ are integration constants
interpreted as the electric and the magnetic charge, respectively, and
the former will be henceforth denoted by $e.$
In what follows we shall confine ourselves to the solutions with
an electric charge only.

 The stress-energy tensor for the line element (\ref{el-gen})
and the Faraday tensor (\ref{F01}) is simply
\begin{equation}
     T_{t}^{t}\,=\,T_{r}^{r}\,=\,- T_{\theta }^{\theta }\,
	 =\,-T_{\phi }^{\phi }\,=\,-\frac{e^{2}}{8\pi\,r^{4}}
    \end{equation}
	                             \label{teep}
and in this form is independent of the metric potentials.

Now we are going to construct approximate solution 
describing static and  spherically-symmetric charged black hole.
The system of differential equations for $m(r)$ and $\psi(r)$
is to be supplemented with the appropriate, physically motivated
boundary conditions.
Our preferred choice, requiring knowledge of the exact location 
of the event horizon, $\rp,$ is 
\begin{equation}
m(\rp)\,=\,\frac{\rp}{2},
                      \label{bound1}
\end{equation}
which, as we shall see, is related to a horizon defined mass of the
black hole. With such a choice we express the solution in terms of the
exact location of the event horizon and the electric charge. On the
other hand it seems natural to express the results in term of the mass
of the system as seen by a distant observer:
\begin{equation}
\lim_{r\to\infty} m(r)\,=\,M_{\infty}.
                        \label{bound2a}   
\end{equation}
For the function $\psi(r)$ we shall 
adopt a natural condition
\begin{equation}
\lim_{r\to\infty}\psi(r)\,=\,0.
                        \label{psi_inf}
\end{equation}

Let us select the boundary conditions of the first type.
Of functions $m(r)$ and $\psi(r)$ we assume that they could be
expanded as 
\begin{equation}
m(r)\,=\,M_{0}(r)\,+\,\sum_{k=1}^{n}\varepsilon^{k} M_{k}(r)\,+
\,{\cal O}(\varepsilon^{n+1})
                                   \label{serM}
\end{equation}
and
\begin{equation}
\psi(r)\,=\,\sum_{k=1}^{n}\varepsilon^{k} \psi_{k}(r)+
\,{\cal O}(\varepsilon^{n+1})
                                   \label{ser_psi}
\end{equation}
It should be noted that $\psi_{0}(r)\,=\,0.$

Since we have assumed the expansion of $m(r)$ 
in the form given by eq. (\ref{serM}),
the condition (\ref{bound1}) could be rewritten in the form:
\begin{eqnarray}
M_{i}(\rp)\,=\,\begin{cases}
           \frac{\rp}{2} & \text{if $i=0$},\\
            0             & \text{if $i\geq 1$}
               \end{cases}
\end{eqnarray}
and  such a choice is a typical mathematical procedure ~\cite{Bender}.
Now, inserting the line element (\ref{el-gen}) into Eqs.~(\ref{grav-sq}),
making use of the expansions (\ref{serM}) and (\ref{ser_psi}), 
and finally collecting the terms
with the like powers of the  parameter $\varepsilon,$ one obtains
the system of the differential equations for $M_{0},$ $M_{k},$ and $\psi_{k}$ 
$(k \geq 1)$ of ascending complexity.
The zeroth-order equations reduce to that of Einstein-Maxwell system,
and, after simple manipulations, lead  to the following integral
\begin{equation}
M_{0}\,=\,{1\over 4}\int r^{2} F dr\,+\,C_{1},
                                     \label{zerothtt}       
\end{equation}
which, when combined with the condition  (\ref{bound1}), yields
\begin{equation}
M_{0}(r)\,=\,{\rp\over 2} + {e^2\over 2\rp} -{e^2\over 2 r}.
                                     \label{M0}
\end{equation}

The first-order equation constructed from the radial component 
of Eq.~(\ref{grav-sq}) 
reads
\begin{eqnarray}
M_{1}^{\prime}&=&\beta\left( \frac{2\,M_{0}^{\prime}}{r^{2}}
-\frac{8\,M_{0}\,M_{0}^{\prime}}{r^{3}}+\frac{2\,{M_{0}^{\prime}}
^{2}}{r^{2}}-\frac{2\,M_{0}^{\prime\prime}}{r}+\frac{5\,M_{0}
\,M_{0}^{\prime\prime}}{r^{2}}-\frac{M_{0}^{\prime}\,M_{0}
^{\prime\prime}}{r}\right. \nonumber \\
&& \left.  +\frac{{M_{0}^{\prime\prime}}^{2}}{2}+M_{0}^{(3)}-\frac
{M_{0}\,M_{0}^{(3)}}{r}-M_{0}^{\prime}\,M_{0}^{(3)}+r\,M_{0}
^{(4)}-2\,M_{0}\,M_{0}^{(4)}\right). 
                                  \label{firsttt}
\end{eqnarray}
On the other hand, the first order equation constructed from the time component
of Eq.~(\ref{grav-sq}) when combined with (\ref{firsttt}),
could be easily integrated to yield
\begin{equation}
\psi_{1}(r)\,=\,\beta\left( M_{0}^{(3)} - {4\over r^{2}}  M'_{0}\right)
\,+\,C_{2},
                                       \label{firstrr}
\end{equation}
where $C_{2}$ is an integration constant.

Inserting Eq.~(\ref{M0}) into (\ref{firsttt}) and integrating the thus
obtained equation one has
\begin{equation}
M_{1}(r)\,=\,\beta\left({\frac {2\,{e}^{2}}{{r}^{3}}}-
\,{\frac {3\,{e}^{2}\rp}{2\,{r}^{4}}}-\,{\frac {3\,{e}^{4}}{2\,\rp\,{r}^{4}}}+
\,{\frac {6\,{e}^{4}}{5\,{r}^{5}}}-\,{\frac {{e}^{2}}{2\,\rp^{3}}}
+\,{\frac {3\,{e}^{4}}{10\,\rp^{5}}}\right).
                                       \label{M1}
\end{equation}
Further, substituting the zeroth-order solution to Eq.~(\ref{firstrr}) 
and making use of the boundary 
condition (\ref{bound1}) gives
\begin{equation}
\psi_{1}(r)\,=\,\beta \frac{e^{2}}{r^{4}}.
                                      \label{psi1}
\end{equation}
Starting from the second-order the differential equations become
more and more complicated and we shall display their solutions only.
After some algebra one has
\begin{eqnarray}
M_{2}(r)&=&\beta^{2}\left({\frac {351}{4}}\,{\frac {{e}^
{4}\rp}{{r}^{8}}}-\,{\frac {36\,{e}^{2}}{{r}^{5}}}\,-\,{\frac {1156\,{e}^{4}}{7\,{r
}^{7}}}\,+\,{\frac {76\,{e}^{2}\rp}{{r}^{6}}}\,+\,{\frac {76\,{e}^{4}}{{r
}^{6}\rp}}\,-\,{\frac {704}{15}}\,{\frac {{e}^{6}}{{r}^{9}}}
\,-\,40\,{
\frac {{e}^{2}\rp^{2}}{{r}^{7}}}\right.
\nonumber \\
&-&\left.40\,{\frac {{e}^{6}}{\rp^{2}{r}^{7}}}\,+\,{
\frac {351}{4}}\,{\frac {{e}^{6}}{\rp\,{r}^{8}}}+\,{3\over 2}{\frac {{e}
^{4}}{\rp^{3}{r}^{4}}}\,-\,{\frac {9}{10}}\,{\frac {{e}^{6}}{{r}^{4
}\rp^{5}}}\,+\,\frac{1}{12}\,{\frac {{e}^{6}}{\rp^{9}}}\,-\,{\frac {3}{
28}}\,{\frac {{e}^{4}}{\rp^{7}}}
\right)
                                       \label{M2}   
\end{eqnarray}
and
\begin{equation}
\psi_{2}\,=\,\beta^{2}\left(32\,{\frac {{e}^{2}\rp}{{r}^{7}}}
\,+\,32\,{\frac {{e}^{4}}{{r}^{7}\rp}}\,
-\,24\,{\frac {{e}^{2}}{{r}^{6}}}\,-\,41\,{\frac {{e}^{4}}{{r}^{8}}
}
 \right),
                                      \label{psi2}
\end{equation}
for the second order, whereas the third order results read
\begin{eqnarray}
M_{3}(r)&=&\beta^{3}\left(
1440\,{\frac {{e}^{2}}{{r}^{7}}}-5508\,{\frac {{e}^{2}\rp}{{r}^{8
}}}+21168\,{\frac {{e}^{4}}{{r}^{9}}}+{\frac {12392664}{385}}\,{\frac 
{{e}^{6}}{{r}^{11}}}+{\frac {331624}{65}}\,{\frac {{e}^{8}}{{r}^{13}}}
+{\frac {2229}{572}}\,{\frac {{e}^{8}}{\rp^{13}}}\right.\nonumber \\
&-&4\,{\frac {{e
}^{2}}{\rp^{7}}}+{\frac {652}{55}}\,{\frac {{e}^{4}}{\rp
^{9}}}-{\frac {2587}{220}}\,{\frac {{e}^{6}}{\rp^{11}}}+6816\,{
\frac {{e}^{6}}{{r}^{9}\rp^{2}}}-76\,{\frac {{e}^{4}}{{r}^{6}\,\rp^{3}}}
-2744\,{\frac {{e}^{2}\rp^{3}}{{r}^{10}}}+{\frac 
{228}{5}}\,{\frac {{e}^{6}}{{r}^{6}\rp^{5}}}\nonumber \\
&-&5508\,{\frac {{e}^
{4}}{\rp\,{r}^{8}}}-{\frac {130732}{5}}\,{\frac {{e}^{4}\rp}{{r}^{10}}}-{\frac {
130732}{5}}\,{\frac {{e}^{6}}{{r}^{10}\rp}}
+{\frac {115176}{11}}
\,{\frac {{e}^{4}\rp^{2}}{{r}^{11}}}-{\frac {51347}{4}}\,{
\frac {{e}^{6}\rp}{{r}^{12}}}\nonumber \\
&+&{\frac {115176}{11}}\,{\frac {{e}^{
8}}{\rp^{2}{r}^{11}}}
-{\frac {51347}{4}}\,{\frac {{e}^{8}}{{
\rp}\,{r}^{12}}}-2744\,{\frac {{e}^{8}}{\rp^{3}{r}^{10}}}+80
\,{\frac {{e}^{4}}{\rp^{2}{r}^{7}}}+32\,{\frac {{e}^{6}}{
\rp^{4}{r}^{7}}}+6816\,{\frac {{e}^{2}\rp^{2}}{{r}^{9}}
}
\nonumber \\
&-&48\left.\,{\frac {{e}^{8}}{\rp^{6}{r}^{7}}}-{\frac 
{351}{4}}\,{\frac {{e}^{6}}{\rp^{3}{r}^{8}}}+{\frac {1053}{20}}
\,{\frac {{e}^{8}}{\rp^{5}{r}^{8}}}-\,{\frac {{e}^{8}}{4\,{r}^{
4}\rp^{9}}}+{\frac {9}{28}}\,{\frac {{e}^{6}}{\rp^{7}{r}
^{4}}}
\right)
                                            \label{M3}
\end{eqnarray}
and
\begin{eqnarray}
\psi_{3}(r)&=&
\beta^{3}\left({\frac {96}{
5}}\,{\frac {{e}^{6}}{{r}^{7}\rp^{5}}}
-3264\,{\frac {{e}^{2}\rp}{{r}^{9}}}-3264\,{\frac {{e}^{4}}{{r}^{
9}\rp}}-32\,{\frac {{e}^{4}}{{r}^{7}\rp^{3}}}-{\frac {74560}{11}}\,{
\frac {{e}^{4}\rp}{{r}^{11}}}-{\frac {74560}{11}}\,{\frac {{e}^{6
}}{{r}^{11}\rp}}\right.
\nonumber \\
&+&\left.2352\,{\frac {{e}^{2}\rp^{2}}{{r}^{10}}}+
{\frac {48384}{5}}\,{\frac {{e}^{4}}{{r}^{10}}}
+2352\,{\frac {{e}^{6}}
{{r}^{10}\rp^{2}}}+{\frac {69572}{15}}\,{\frac {{e}^{6}}{{r}^{
12}}}+1080\,{\frac {{e}^{2}}{{r}^{8}}}\right).
                                           \label{psi3}
\end{eqnarray}

It is possible to express this result in a more familiar form
introducing the horizon defined mass $M,$ i. e. to represent
solution in terms of $(e,\,M)$ rather than $(e,\,\rp).$
It could be easily 
done employing equality
\begin{equation}
M\,=\,{\rp\over 2}\,+\,{e^{2}\over 2 \rp}
                                                \label{bhmass}
\end{equation}
and repeating calculations with
\begin{equation}
M_{0}(r)\,=\,M\,-\,\frac{e^{2}}{2r}.
\end{equation}
Now the {\it exact} location of the event horizon is
related to the horizon defined mass by a classical formula 
\begin{equation}
r_{+}\,=\,M\,+\,(M^{2} -e ^{2})^{1/2}.
\end{equation}
It should be noted, however, that although the zeroth-order equation
gives the exact location of the event horizon, the same is not true 
for its second root, 
\begin{equation}
r_{c}\,=\,\frac{e^{2}}{\rp}.
\end{equation}
 Indeed, it could be shown that the inner horizon, $r_{-},$ is given by
 \begin{eqnarray}
 r_{-}&=&r_{c}\,+\,\frac{\beta}{5}\left(\frac{3e^{4}}{\rp^{5}}\,-\,\frac{2e^{2}}{\rp^{3}}
 \,-\,\frac{2}{\rp}\,-\,\frac{2\rp}{e^{2}}\,+\,\frac{3\rp^{3}}{e^{4}}\right)
 \,+\,\beta^{2}\left(\,{\frac {{e}^{6}}{6\rp^{9}}}-
 {\frac {214}{525}}\,{\frac {{e}^{4}}{\rp^{7}}}
 \right.
 \nonumber \\
 &&+\left.{\frac {38}{525}}\,\frac {{e}^{2}}{\rp^{5}}\,+\,
 {\frac {206}{525\rp^{3}}}\,-{\frac {131}{105{{e}^{2}\rp}}   }\,+\,
 {\frac {58}{105}}\,{\frac \rp{{e}^{4}}}+{\frac {374}{525}}\,{\frac {\rp^{3}}{{e}^{6}}}
 +{\frac {542}{525}}\,{\frac {\rp^{5}}{{e}^{8}}}
 -{\frac {191}{150}}\,{\frac {\rp^{7}}{{e}^{10}}}\right)\,+\,{\cal O}(\beta^{3}), \nonumber \\
\end{eqnarray}
and, moreover, simple calculation shows that for $\rp = |e|$ both
horizons coincide to required order. Such a behaviour strongly
suggests that this very relation describes degenerate  horizon of the
extreme black hole. Since the extreme black holes deserve more
accurate treatment we shall postpone further analysis of this and
related problems to Sec.~\ref{sec4}.

\section{Temperature and entropy}
                   \label{sec3}
One of the most important characteristics of the black hole is its
Hawking temperature, $T_{H}.$ To investigate how $T_{H}$ is modified by
quadratic terms, we employ the standard procedure of the Wick
rotation. The thus obtained complexified line element
has no conical singularity as $r \to \rp$ provided the time coordinate
is periodic with a period $P,$ which is to be identified with the
reciprocal of $T_{H}.$ Elementary considerations carried out for the
general  static and spherically symmetric spacetime describing a black hole 
lead to the formula
\begin{equation}
P\,=\,\frac{1}{T_{H}}\,=\,4\pi \lim_{r\to \rp}(g_{tt} g_{rr})^{1/2}
\left(\frac{d}{dr}g_{tt} \right)^{-1}.
                                 \label{period}
\end{equation}
For the line element (\ref{el-gen}) and the boundary condition
(\ref{bound1}) the above result assumes simple form
\begin{equation}
T_{H}\,=\,\frac{e^{\psi(\rp)}}{2\pi\rp^{2}}\left(\frac{\rp}{2}\,
-\,\rp\frac{dm(r)}{dr}_{|r=\rp}\right).
                                 \label{tem1}  
\end{equation}
Making use of Eqs.~(\ref{M0}) and (\ref{M1}--\ref{psi3}), and
collecting the terms with the like powers of $\beta$, after massive simplifications,  
one obtains
\begin{eqnarray}
T_{H}&=&\frac{1}{4\pi\rp}\left(1\,-\,\frac{e^{2}}{\rp^{2}} \right)\,+\,
\frac{\beta e^{2}}{4\pi \rp^{5}}\left(1\,-\,\frac{e^{2}}{\rp^{2}} \right)\,-\,
\frac{\beta^{2} e^{4}}{8\pi\rp^{9}}\left(1\,-\,\frac{e^{2}}{\rp^{2}} \right)\nonumber \\
&&+\frac{\beta^{3}e^{2}}{440\pi \rp^{13}}\left(1\,
-\,\frac{e^{2}}{\rp^{2}} \right)\left(880\rp^{4}\,-\,3232 e^{2}
\rp^{2}\,+\,2455 e^{4}\right)\,+\,{\cal O}(\beta^{4}).
                                      \label{temp2}
\end{eqnarray}
We suspect that the common factor in the above expression appears in
all higher-order terms and for $\rp\,=\,|e|$ the black hole
temperature approaches zero. We shall return to this problem later in
the text.

We have not, as yet, imposed any constraints on the parameter $\beta,$
but now we are going to examine the consequences of the non-tachyon
condition, which, in the case at hand, is simply
$
\beta \,<\,0.
$
For the nonextremal black hole the terms proportional to $\beta$ and
$\beta^{2}$ in the right hand side of the equation (\ref{temp2}) are
strictly negative, and, therefore, one concludes that the temperature
(to this order)  is lower as compared with the Reissner-Nordstr\"om
black hole described by the same values of $\rp$ and $e.$ On the other
hand the higher-order terms proportional to $\beta^{k},$ (for $k\geq
3$) change sign, however, for reasonable values of the coupling
constant their contribution to a total temperature is negligible.

The entropy of the black hole in quadratic gravity may be  calculated
using various methods. Here we shall employ the  Wald's Noether charge
technique~\cite{Wald:1993nt}, which, as has been shown in Refs.~\cite{Myers1,Ted}, may be
safely applied for Lagrangians of the type (\ref{actionf})  and leads
to remarkably simple and elegant general result
\begin{equation}
S\,=\,\frac{1}{4}\int_{\Sigma}\,d^{2}x\sqrt{h}
\left[1\,+\,2\alpha R\,+\,\beta\left(R\,-\,h^{ab} R_{ab} \right) \right],
\end{equation}
where $h_{ab}$ is induced metric on $\Sigma$ and the surface integral is taken 
across arbitrary section of the event horizon. In our representation, the entropy,
after some algebra
may be compactly written as
\begin{eqnarray}
S&=&{A\over 4}\,-\,\frac{8\pi^{2} \beta}{A} e^{2}\,
-\,{1024\pi^{4}e^{2}\beta^{3}\over A^{5}}\left(A - 4\pi e^{2}\right)^{2}\nonumber\\
&&+\,\frac{16384 \pi^{5}e^{2}\beta^{4}}{A^{7}}\left(A - 4\pi e^{2}\right)^{2}
\left(26\pi e^{2} - 5A\right)
\,+\,{\cal O}(\beta^{5}),
                              \label{entr}
\end{eqnarray}
where $A\,=\,4\pi\rp^{2}$ is the surface of the event horizon.
We attribute discrepancies between our result and the entropy 
computed by CLA  to errors in their metric tensor.

\section{The extremal configuration and its near horizon geometry}
                                        \label{sec4}
In this section we shall investigate the important issue of the extremal
black holes. On general grounds one expects that the extremal 
configuration is the one in which (at least) two horizons merge.  
It means that in the simplest case the first two terms of the expansion 
\begin{equation}
f(r)\,=\,f(\rp)\,+\,\frac{df}{dr}_{|r=\rp}(r-\rp)\,
+\,\frac{1}{2}\frac{d^{2}f}{dr^{2}}_{|r=\rp}
(r-\rp)^{2}\,+\,...
                   \label{expansion}
\end{equation}
vanish. 
It is evident that if first $n$ terms in the above expansion are absent
one has a $n-$fold merging of $n$ horizons.
In a view of the further applications we also expect 
regularity of the function $\psi(r)$
as the degenerate event horizon is approached:
\begin{equation}
|\psi (\rp)| < \infty,\hspace{1cm} |\frac{d\psi(r)}{dr}_{|r=\rp} | <\infty .
                                             \label{reg1a}
\end{equation}

In order to determine the location of the event horizon, i. e. to
relate integration constants $\rp$ and $e$ one has to consider
consequences of vanishing of the black hole temperature (surface
gravity). Eq.~(\ref{temp2}) strongly indicates that $\rp = |e|.$ To
demonstrate validity of  this relation let us consider the $(rr)$
component of (\ref{grav-sq}). It could be shown that if $f'(\rp) = 0,$
the conditions (\ref{reg1a}) are satisfied, and, additionally,
\begin{equation}
|f^{(3)}(\rp)|\,<\,\infty, \hspace{1cm} |\psi^{(3)}(\rp)|\,<\,\infty,
\end{equation}
then the second derivative of  $f$ computed at $r\,=\,\rp$ 
satisfies the constraint equation
\begin{equation}
\frac{1}{\rp^{2}}\,+\,\frac{\beta}{\rp^{4}}\,-\,
\frac{\beta}{4}\left(\frac{d^{2}f}{dr^{2}}_{|\rp}\right)^{2}\,
=\,\frac{e^{2}}{\rp^{4}}.
                                \label{constr1}
\end{equation}
The metric of the electrically charged black hole is to be independent
of the coupling constant $\alpha$~\cite{Whitt:1984pd,Campanelli:1994sj}.
Repeating calculations for the
$(rr)$ component of the tensor $\phantom{}^{(1)}H_{a}^{b}$ at
$r\,=\,\rp$ one has
\begin{equation}
\phantom{}^{(1)}H_{r}^{r}(\rp)\,=\,\frac{2}{\rp^{4}}\,-
\,\frac{1}{2}\left(\frac{d^{2}f}{dr^{2}}_{|\rp}\right)^{2}.
                                    \label{H_a}
\end{equation}
Absence of  terms proportional to  
$\alpha^{k}$ $(k=0,1,2...)$ 
requires $\phantom{}^{(1)}H_{a}^{b}$ to vanish identically, and, since
the both conditions are to be satisfied simultaneously one obtains
expected exact result:
\begin{equation}
\rp\,=\,|e|.
                                  \label{extreme1}
\end{equation}
Equivalently, expressing the location of $\rp$ in terms of the horizon
defined mass yields
\begin{equation}
\rp\,=\,M.
                                 \label{extreme2}
\end{equation}
Obtained relations are identical with those which characterize
geometry of the extremal Reissner-Nordstr\"om. As the particular form
of the traceless stress-energy tensor played the decisive role in our
derivation one should not expect simple generalizations. In a more
general case, as for example that of the nonlinear electrodynamics,
there is an explicit dependence on the coupling constant $\alpha,$ and
the stress-energy tensor lacks simplicity of it Maxwellian analog.
Indeed, repeating calculations for a line element of the type
(\ref{el-gen}) and an arbitrary stress-energy tensor, one
has
\begin{equation}
  \left(\beta\,+\,2\alpha\right)\left[\left(\frac{1}{2}f''(\rp)\right)^{2}\,
  -\,\frac{1}{\rp^{4}}\right]\,-
  \,\frac{1}{\rp^{2}}\,=\,8\pi T_{r}^{r}(\rp)
                                      \label{ehr}
\end{equation}
and
\begin{equation}
-\left(\beta\,+\,2\alpha\right)\left[\left(\frac{1}{2}f''(\rp)\right)^{2}\,
  -\,\frac{1}{\rp^{4}}\right]\,+
  \,\frac{1}{2}f''(\rp)\,=\,8\pi T_{\theta}^{\theta}(\rp).
                                      \label{ehtet}  
\end{equation}
It should be noted that in order to obtain Eq. (\ref{ehtet}) one has to assume
$|f^{(4)}(\rp)|<\infty$ and $|\psi^{(4)}(\rp)|<\infty.$
The spherical symmetry and the field equations 
at the event horizon respectively give
\begin{equation}
T_{\theta}^{\theta}\,=\,T_{\phi}^{\phi}, \hspace{1cm} T_{t}^{t}\,=\,T_{r}^{r}. 
\end{equation}
Combining the above equations one obtains
\begin{equation}
f''(\rp)\,-\,\frac{2}{\rp}\,=\,8\pi T_{a}^{a}(\rp),
                                      \label{ftrace}
\end{equation}
where $T_{a}^{a}$ is the trace of the stress-energy tensor.
\par
As is well-known the closest vicinity of the event horizon of the
extremal Reissner-Nordstr\"om black hole, after the coordinate
transformation
\begin{equation}
r\,=\,\rp\left(1 + \frac{\rp}{y} \right),
\end{equation}
could be approximated by the Bertotti-Robinson line
element~\cite{bertotti,robinson}
\begin{equation}
ds^{2} = {\frac{\rp^{2} }{y^{2}}} \left( - dt^{2} + dy^{2} + y^{2}
d\Omega^{2}\right).
                       \label{BR}
\end{equation}
Employing new static coordinates the Bertotti-Robinson 
line element can be rewritten 
in the form 
\begin{equation}
ds^{2}\,=\,\rp^{2}\left( -\sinh^{2}\chi d\tau^{2}\,+\,d\chi^{2}\,+\,d\theta^{2}\,+\,
\sin^{2}\theta d\phi^{2}\right)
\end{equation}
or 
\begin{equation}
ds^{2}\,=\,-\left(\frac{x^{2}}{\rp^{2}} -1\right)dT^{2}\,+
\,\left(\frac{x^{2}}{\rp^{2}} -1\right)^{-1}
dx^{2}\,+\,\rp^{2}\left(d\theta^{2}\,+\,
\sin^{2}\theta d\phi^{2} \right).
\end{equation}
The latter form is particularly useful in demonstrating that topology
of the Bertotti-Robinson solution is ${\rm AdS}_{2}\times {\rm
S}^{2},$ i.e., it is a simple topological product of a
(1+1)-dimensional anti-deSitter spacetime and two-sphere of radius
$\rp.$

One expects that this spacetime also plays some role for the extremal
black holes in the quadratic gravity. Indeed, first observe that
because of simplicity of the Bertotti-Robinson line element  the
tensors $^{(1)}H_{ab}$ and $^{(2)}H_{ab}$ as well as the curvature
scalar vanish. The only contribution to the left hand side of
(\ref{grav-sq}) comes, therefore, from the Ricci tensor, which for the
metric (\ref{BR}) is given by
\begin{equation}
R_{t}^{t}\,=\,R_{r}^{r}\,=\,-R_{\theta}^{\theta}\,=
\,-R_{\phi}^{\phi}\,=\,-\frac{1}{\rp^{2}}.
\end{equation}
As the stress-energy tensor of the electromagnetic field
is simply
\begin{equation}
T_{t}^{t}\,=\,T_{r}^{r}\,=\,-T_{\theta}^{\theta}\,=
\,-T_{\phi}^{\phi}\,=\,-\frac{1}{8\pi\rp^{2}},
\end{equation}
one concludes that the line element (\ref{BR}) is an exact solution if
$\rp\,=\,|e|.$ Moreover, inspection of (\ref{grav-sq}) suggests that,
after redefinition of the time coordinate, the dominant contribution
to the metric potentials $g_{tt}$ and $g_{rr}$ near the event horizon
comes from the function $M_{0}(r).$ It follows then that locally it
resembles the geometry of the extremal Reissner-Nordstr\"om black
hole.

To investigate the role played by the Bertotti-Robinson solution in
more details let us return to the expansion (\ref{expansion}) and ask
when the near horizon geometry of the extremal black hole is  that
described by the line element (\ref{BR}). In the vicinity of the event
horizon the line element may be written as
\begin{equation}
ds^{2}\,=\,- e^{2\psi (r_{+})} F(r-r_{+})^{2}dt^{2} + 
{\frac{1}{F (r - r_{+})^{2}}} dr^{2} + r_{+}^{2} d\Omega^{2},
                                      \label{lineel_ser}
\end{equation}
where 
\begin{equation}
F ={\frac{1}{2}}{\frac{d^{2} f}{d r^{2}}}_{|r=r_{+}}.
\end{equation}
It could be easily demonstrated  that expressing the line element
in terms of a new coordinate $y$ defined by means of the 
relation
\begin{equation}
r = r_{+} \left( 1 + {\frac{r_{+}}{e^{\psi (r_{+})} y }}\right),
\end{equation}
one obtains the line element (\ref{BR}) provided 
\begin{equation}
\frac{d^{2}f}{dr^{2}}_{|r=\rp}\,=\,\frac{2}{\rp^{2}}.
                                    \label{seconDa}
\end{equation}
Inspection of (\ref{H_a}) shows that it is precisely the relation
which are satisfied by extremal black hole.
It should be noted that by (\ref{ftrace}) our demonstration requires 
the trace of the stress-energy tensor to vanish at $r\,=\rp$
only.

On the other hand, if $f''(\rp)$ does not obey Eq.~(\ref{seconDa}), the topology
of the solution (\ref{lineel_ser}) is still a simple 
product of ${\rm AdS}_{2}\times {\rm S}^{2},$
but with different modulus of curvatures. Indeed, it could be easily shown that
\begin{equation}
R\,=\,K_{{\rm AdS}_{2}}\,+\,K_{{\rm S}^{2}},
\end{equation}
where
\begin{equation}
K_{{\rm AdS}_{2}}\,=\,-2 F \hspace{1cm}{\rm and}
\hspace{1cm}K_{{\rm S}^{2}}\,=\,\frac{2}{\rp^{2}}.
\end{equation}
Upon substitution
\begin{equation}
r\,=\,\rp\,+\,\frac{1}{e^{\psi(\rp)}F\,y}
\end{equation}
one obtains
\begin{equation}
ds^{2}\,=\,\frac{1}{Fy^{2}}\left( -dt^{2}\,+\,dy^{2}\right)\,
+\,\rp^{2}\left(d\theta^{2}\,+\,
\sin^{2}\theta d\phi^{2}\right).
                          \label{br_gen}
\end{equation}
Finally observe that vanishing of the curvature scalar as $r\,\to \,\rp$
yields (\ref{seconDa}), and the line element (\ref{br_gen}) reduces to (\ref{BR}).

\section{A distant observer point of view}
                               \label{sec5}

In this section we shall briefly examine consequences of the second
choice of the boundary conditions as given by Eqs.(\ref{bound2a}) and (\ref{psi_inf}). This problem has
been considered earlier by CLA. Unfortunately, their metric tensor and
hence other characteristics of the charged black holes in quadratic
gravity contain errors in the terms proportional to $\beta^{3}.$

Before proceeding further  let us return to the function
$m(r)$ constructed with the aid of the condition (\ref{bound1}). Its
limit as $r \to \infty$ is simply
\begin{eqnarray}
\Mi&=&\frac{\rp}{2}\,+\,\frac{e^{2}}{2\rp}\,+
\,\beta\left(\frac{3e^{4}}{10\rp^{5}}\,
-\,\frac{e^{2}}{2\rp^{3}} \right)
\,+\,\beta^{2}\left(\frac{e^{6}}{12\rp^{9}}\,-
\,\frac{3e^{4}}{28\rp^{7}} \right)\nonumber \\
&&-\beta^{3}\left(\frac{4e^{2}}{\rp^{7}}\,-\,\frac{652e^{4}}{55\rp^{9}}\,
+\,\frac{2587e^{6}}{220\rp^{11}}\,-
\,\frac{2229e^{8}}{572\rp^{13}} \right)\,+\,{\cal O}(\beta^{4}),
                                       \label{distant_mass}
\end{eqnarray}
and is interpreted as a total mass of a system as seen by a distant
observer and expressed in terms of the exact location of the event
horizon and electric charge. On the other hand one can represent
solution to the system (\ref{grav-sq}) in terms of the total mass as
seen from large distances and the electric charge from the very
beginning. In this case the boundary conditions for the expansion
(\ref{serM}) are to be rewirtten  in the form:
\begin{eqnarray}
\lim_{r\to \infty}M_{i}(r)\,=\,\begin{cases}
           \Mi & \text{if $i=0$},\\
            0             & \text{if $i\geq 1$},
               \end{cases}
\end{eqnarray}
while the condition (\ref{psi_inf}) remains, of course, intact.
Repeating calculations order by order with the new boundary conditions
one obtains
\begin{eqnarray}
&&m(r)\,=\,M_{\infty }-{\frac{{e}^{2}}{2r}}+\beta \,\left( 2\,{\frac{{e}^{2}}{{r
}^{3}}}-3\,{\frac{{e}^{2}M_{\infty }}{{r}^{4}}}+\,{\frac{6\,{e}^{4}}{5{r}^{5}
}}\right)\nonumber  \\
&&+{\beta }^{2}\left( 152\,{\frac{{e}^{2}M_{\infty }}{{r}^{6}}}-{\frac{596}{7
}}\,{\frac{{e}^{4}}{{r}^{7}}}-{\frac{704}{15}}\,{\frac{{e}^{6}}{{r}^{9}}}
-160\,{\frac{{e}^{2}{M_{\infty }}^{2}}{{r}^{7}}}+{\frac{351}{2}}\,{\frac{{e}
^{4}M_{\infty }}{{r}^{8}}}-36\,{\frac{{e}^{2}}{{r}^{5}}}\right)\nonumber  \\
&&+{\beta }^{3}\left( 27264\,{\frac{{e}^{2}{M_{\infty }}^{2}}{{r}^{9}}}
-11016\,{\frac{{e}^{2}M_{\infty }}{{r}^{8}}}+1440\,{\frac{{e}^{2}}{{r}^{7}}}
+{\frac{4330344}{385}}\,{\frac{{e}^{6}}{{r}
^{11}}}-{\frac{179144}{5}}\,{\frac{{e}^{4}M_{\infty }}{{r}^{10}}}\right.  \nonumber\\
&&\left. +{\frac{
331624}{65}}\,{\frac{{e}^{8}}{{r}^{13}}}+{\frac{460704}{11}}\,{\frac{{e}^{4}{
M_{\infty }}^{2}}{{r}^{11}}}-{\frac{51347}{2}}\,{\frac{{e}^{6}M_{\infty }}{{r
}^{12}}}-21952\,{\frac{{e}^{2}{M_{\infty }}^{3}}{{r}^{10}}}
+7536\,{\frac{{e}^{4}}{{r}^{9}}}\right)\,+\,{\cal O}(\beta^{4}) 
\nonumber\\
                            \label{minf}
\end{eqnarray}
and
\begin{eqnarray}
\psi(r)\,&=&\,\beta\frac{e^{2}}{r^{4}}\,-\,\beta^{2}\left(\frac{24e^{2}}{r^{6}}\,
-\,\frac{64 e^{2}\Mi }{r^{7}}\,+\,\frac{41 e^{4}}{r^{8}} \right)\nonumber\\
&&+\,\beta^{3}\left(\frac{24864 e^{4}}{5r^{10}}\,+\,\frac{9408e^{2} \Mi^{2}}{r^{10}}\,+
\,\frac{69572e^{6}}{15r^{12}}\right.\nonumber \\
&&\left.+\,\frac{1080e^{2}}{r^{7}}\,
-\,\frac{6528e^{2}\Mi }{r^{9}}\,-\,\frac{149120e^{4} \Mi }{11r^{11}}\right)\,
+\,{\cal O}(\beta^{4}).
\nonumber \\
                           \label{psiinf}
\end{eqnarray}
Eqs.~(\ref{minf}) and (\ref{psiinf}) are sufficient to determine 
$g_{tt}$ and $g_{rr}$ to ${\cal O}(\beta^{4}).$
The thus obtained metric tensor
differs from that obtained by CLA.
To demonstrate that our calculations lead to correct
results let us make use of the consistency check.
Inserting Eq.~(\ref{distant_mass}) into (\ref{minf}) and (\ref{psiinf}) 
one obtains, as expected, Eqs.~(\ref{M0}) and (\ref{M1}--\ref{psi3}).
Since the calculations have been carried out independently we 
conclude that our results are correct.

To determine location of the event horizon
one can either iteratively solve the equation
\begin{equation}
g_{tt}(\rp)\,=\,0
\end{equation}
or revert relation (\ref{distant_mass}). Both methods give the same result, which reads 
\begin{eqnarray}
\rp\,&=&\,r_{0}\,+\,\beta\,\frac{e^{2}\,\left(5 r_{0}^{2} -
3 e^{2} \right)}{5r_{0}^{3}\left(r_{0}^{2}-e^{2} \right)}\,-\,
\beta^{2}\,\frac{e^{4}\left(2925 r_{0}^{6} - 6515 r_{0}^{4}e^{2} + 
5095 r_{0}^{2}e^{4} - 1337 e^{6}\right)}
{1050 r_{0}^{7}\left(r_{0}^{2} -e^{2} \right)^{3}}
\nonumber \\
&&-\,\beta^{3}\frac{e^{2}}{\left(r_{0}^{2} -e^{2} \right)^{5}}
\left({\frac {17268913}{75075}}\,{\frac {{e}^{10}}{r_{0}^{7}}}
\,-\,{\frac {61234643}{750750}}\,{\frac {{e}^{12}}{r_{0}^{9}}}
\,-\,{\frac {137993}{770}}\,{\frac {{e}^{4}}{r_{0}}}
\right. \nonumber \\
&&\left.+
{\frac {3265043}{10010}}{\frac {{e}^{6}}{r_{0}^{3}}}
\,-\,{\frac {26686967}{75075}}\,{\frac {{e}^{8}}{r_{0}^{5}}}
\,+\,{\frac {436497}{35750}}\,{\frac {{e}^{14}}{r_{0}^{11}}}
\,-\,8\,r_{0}^{3}-{\frac {3064}{55}}\,{e}^{2}r_{0}\right)\,+\,
{\cal O}(\beta^{4}),\nonumber \\
\end{eqnarray}
where
\begin{equation}
r_{0}\,=\,\Mi\,+\,\left(\Mi^{2} -e^{2}\right)^{1/2}.
\end{equation}

Repeating the steps of Sec.~\ref{sec3} necessary to compute the Hawking temperature,
and expressing the final result in terms of $e$ and $r_{0},$
one has
\begin{eqnarray}
T_{H}&=&\frac{1}{4\pi r_{0}^{3}}\left(r_{0}^{2} - e^{2}\right)\,
+\,\beta\,\frac{e^{4}\left(2 r_{0}^{2} 
- e^{2}\right)}{5r_{0}^{7}\left(r_{0}^{2} - e^{2}\right)}\nonumber \\
&&+\,\beta^{2}\frac{e^{4}}{\pi r_{0}^{11}\left(r_{0}^{2} - e^{2}\right)^{3}}
\left({\frac {8}{25}}\,{e}^{8}-{\frac {4}{21}}\,{e}^{2} r_{0}^{6}+{
\frac {113}{75}}\,{e}^{4} r_{0}^{4}-{\frac {676}{525}}\,{e}^{6}
 r_{0}^{2}-\frac{3}{7}\, r_{0}^{8}\right)\,+\,{\cal O}(\beta^{3}).
\nonumber \\
\end{eqnarray}
Although the extremal configuration can be studied in $(e,\,\Mi )$
representation it is not the best choice. Indeed, simple calculations
give for the location of the event horizon
\begin{equation}
r_{+}\,=\,M_{\infty}\,+\,\frac{1}{5\Mi}\beta\,
-\,\frac{17}{1050 M_{\infty}^{3}}\beta^{2}\,+\,\frac{317}{68250\Mi^{5}}\beta^{3}
\,+\,{\cal O}(\beta^{4}),
\end{equation}
whereas relation between the total mass as seen 
by a distant observer and the electric charge has the form:
\begin{equation}
M^{2}_{\infty}\,=\,e^{2}\,-\,\frac{2}{5}\beta\,-\,\frac{4}{525\,e^{2}}\beta^{2}\,-\,
\frac{8}{1365e^{4}}\beta^{3}\,+\,{\cal O}(\beta^{4}).
\end{equation}
This could be contrasted with the simple relation between $\rp,$ and $|e|$
given by Eqs.~(\ref{extreme1}) and~(\ref{extreme2}).
Finally observe, that since the entropy as given by Eq. (\ref{entr}) is expressed 
in terms of the surface of the event horizon  
and the electric charge, we expect that it should be described by the same 
formula for both choices of the conditions
(\ref{bound1}) and (\ref{bound2a}).
\section{Conclusion and summary}
                \label{sec6}
				
In this paper we have constructed iterative solutions describing
spherically-symmetric and static black holes to the equations of the
fourth-order gravity with the source term given by the stress-energy
tensor of the electromagnetic field . Obtained line elements are
parametrized by two integration constants which are related to the
electric charge and the exact location of the event horizon. The thus
computed metric potentials enabled construction of the basic
characteristics of the black hole: its Hawking temperature and
entropy. 

Special emphasis has been  put on the extremal black holes.
Specifically,  it has been explicitly demonstrated that in the
extremal limit, the exact location of the (degenerate) event horizon
is given by $\rp\,=\,|e|.$ It was shown that, similarly to the
classical Reissner-Nordstr\"om solution, the near-horizon geometry of
the charged black holes in quadratic gravity, when expanded into the
whole manifold, is simply that of Bertotti and Robinson. As a byproduct
of our investigations we obtained a simple equation that relate
horizon value $f''$ with the trace of the stress-energy tensor, which,
for solutions of the equations of quadratic gravity describing black
holes, may serve as an useful criterion for possessing the near-horizon
geometry of the Bertotti-Robinson type. 

Similar considerations have been carried out for the boundary
conditions of second type which employ the electric charge and the  mass of the
system as seen by a distant observer. 
Moreover, it has been explicitly demonstrated how to relate 
appropriate results obtained within the framework of each method.

Returning to the extremal black holes we observe that there are good
reasons to believe that they are qualitatively different from the
nonextremal ones. A proper distance, for example, between two points,
one of which resides on the event horizon, is infinite. It could be
easily seen from the integral
\begin{equation} 
\int \sqrt{g_{rr}}dr \,\sim\,\frac{1}{\rp}\ln(r-\rp),
\end{equation} 
as it diverges in the limit $r \to \rp.$ Moreover, the entropy
remains nonzero as $A\to 4\pi \rp^{2},$ and depends on the electric
charge. This behaviour  clearly violates the Nernst formulation of the
third law of thermodynamics, which states that the entropy of a system
must go to zero or a universal constant as $T\to 0.$ On the other
hand, even if the entropy of the extremal black hole
vanish~\cite{Hawking} there are still problems simply because of its
noncontinous nature. Indeed, zero is not the limit to which the
entropy of the nearly extremal black holes tends. One can argue that
this behaviour should not be treated as to worrisome simply because
the Nernst formulation probably should not be considered as the
fundamental law of thermodynamics. For a recent discussion of this
issue see~\cite{Wald:1997qp}. On the other hand according to more
radical opinions this failure  indicates that the extremal black holes
must not be treated as thermodynamic systems to which one can assign
the notion of temperature and entropy~\cite{Tony}. Of course,  we are
unable here to judge which option should be treated seriously. All we
can say now is that the entropy as given by (\ref{entr}) (probably)
could be safely used for nonextremal black holes. However, the
particular case of the extreme black holes certainly deserves further
study.

Finally, let us observe that the methods of this paper could by easily
generalized to other sources, such as, for example, nonlinear
electrodynamics. This group of problems is under active
investigations and the results will be published elsewhere.



\end{document}